# Study on Dynamic Matching and Dynamic Characteristics of Hydrostatic Transmission System of Forklift Truck

AN Ying[1]   YI Ge[1]   LIU Guoliang[2]   SUN Rongwu[2]   ZHAO Junbo[2]
WANG Xiaolan[2]   GAO Shuai[1]

(1. College of Mechanical and Electrical Engineering, Hunan University of Science
and Technology, Xiangtan Hunan 411201, China;
2. Hunan Sinoboom Intelligent Equipment Co.,ltd.,Changsha Hunan 410600,China)

**Abstract**：In the fields of agricultural machinery, construction equipment, and special-purpose vehicles, hydrostatic transmission (HST) drive systems have witnessed a significant increase in market penetration. With the rapid development of intelligent and environmentally friendly trends, higher requirements have been imposed on HST control systems regarding operational efficiency, control accuracy, and adaptability to working conditions. This study focuses on a forklift truck drive system equipped with a dual variable-displacement pump-motor assembly. A mathematical model of the HST system was constructed, with detailed analysis of each stage during vehicle acceleration. A power matching strategy was proposed to enhance dynamic performance, whose core concept consisted of three elements: 1) correlating the driver's pedal signal with the target engine speed of the diesel engine, 2) dynamically adjusting motor displacement to maximize acceleration capability, and 3) coordinating pump displacement variations. Furthermore, a dynamic compensation mechanism for control variables was introduced to improve system responsiveness. The controller was evaluated under multiple operating conditions, with results demonstrating superior acceleration performance, rapid response, and precise control characteristics. These results validate the effectiveness of the proposed control strategy and its robust load adaptability in hydrostatic drive systems.

**Keywords**：Hydrostatic Transmission; Dual-variable-displacement; Dynamic Matching；Dynamic Compensation Adjustment；

## 0 Introduction

HST system composed of pumps and motors is one of the main transmission methods for heavy-duty vehicles[1-4], which has the load adaptability and stepless regulation of transmission ratio, and the power output is more stable than mechanical method[5-7]. This makes them particularly well-suited for engineering vehicles that experience significant load variations[8-12] leading to their widespread utilization in this sector[13-15].

This study takes a forklift truck as the research subject and conducts a comprehensive investigation into the power matching strategy and performance of its HST system. As shown in Fig.1, the HST system consist of a variable displacement pump, variable displacement motor, two-speed transmission, and wheel-end reducer. The hydraulic system employs volumetric speed control, which achieves stepless transmission ratio regulation by adjusting the displacements of the variable pump and motor[16,17].

Current research on HST predominantly focuses on those utilizing single-variable system. To enhance vehicle off-road capability, Reference[18] introduced an auxiliary hydrostatic drive system in conventional two-wheel-drive configurations, integrating a variable-displacement pump with a famous with a fixed-displacement motor. This system dynamically adjusts pump displacement through PID control while responding to load demands, enabling adaptive operation across diverse road conditions. Reference[19] proposed a constant-speed control methodology for fixed-displacement pump and variable-displacement motor system, achieving stable constant-speed operation under constant-flow conditions. The study systematically investigated the interrelationship between the dynamic response characteristics of the motor and parameters of pump, revealing critical coupling mechanisms during transient operations. Reference(4) developed a leakage detection framework for variable-displacement pump-fixed-displacement motor systems, utilizing virtual test environments to systematically analyze the effects of leakage points and flow on hydraulic performance. This study further proposed a pump health assessment methodology based on flow characteristic monitoring. Reference[20] developed a HST system model for variable-displacement pump coupled with fixed-displacement dual-motors. The model's validity was verified through comparative analysis with experimental dataset, followed by comprehensive parametric estimation. The substantial agreement between simulated results and experimental measurements demonstrates the model's fidelity, enabling its integration with control strategies for advanced performance analysis of hydraulic drivetrains. Reference[21] developed a fully parametric model for floating- distributor plate piston pump/motor systems, systematically investigating the dynamic pressure evolution in hydraulic circuits. This microscopic-level analysis elucidated intrinsic system behaviors, offering guidelines for engineering application. Reference[22] investigated a HST system composed of variable-displacement pump and fixed-displacement motor. Through experimental data analysis, parameters were identified, and a comprehensive efficiency model for both hydraulic pump and motor was established. Reference[23] implemented a fuzzy-PID composite control system in HST with variable-displacement pump and fixed-displacement motor, which improved the matching performance of the system. Reference[24] investigated walking systems of harvesting machinery equipped with variable-displacement pump and fixed-displacement motor, proposing a cross-coupled fuzzy-PID control strategy that resolved stability deficiencies. Reference[25] proposed a dual-feedforward with fuzzy-PID composite strategy for mechanical-hydraulic hybrid system. By implementing constant speed control of motor, the study resolved output stability issues in the drivetrain under varying operational conditions. Reference(3) proposed a multivariable control strategy based on fuzzy sliding mode control, and the control effects of common control strategies are compared. The simulation results show that the proposed control strategy and system have good matching and can adapt to disturbances actively. Field experiment results show that the proposed solution can achieve smooth automatic shifting of wheel loaders, while effectively adjusting system pressure, flow, and power.

Integrating the aforementioned research results, current investigations into HST systems have addressed performance across levels ranging from system-level to micro-scale level. This comprehensive research framework has established a theoretical foundation for future research and engineering applications in this field. However, with demands for enhanced equipment performance, dual-variable configurations with variable-displacement pumps and motors are increasingly prevalent. While such architectures provide more degrees of freedom in power matching through pump/motor decoupling, they simultaneously introduce heightened complexity in power matching. Existing research on dual-variable systems remains limited in addressing these challenges.

The dual-variable HST system for forklift trucks studied in this paper. Based on the structural characteristics and combined with operational requirements, this paper formulates the power matching strategy and variable control scheme. Adaptive compensation mechanisms are implemented under specific operating conditions to refine system performance. Comprehensive experimental validation confirms the efficacy of the proposed control architecture in achieving desired operational objectives.

# 1 Dual-variable hydrostatic drive system model development

The study focuses on a forklift truck equipped with a dual-variable HST system, where the variable-displacement pump and variable-displacement motor form a pressure-adaptive closed-loop circuit[29], as shown in Fig.1"

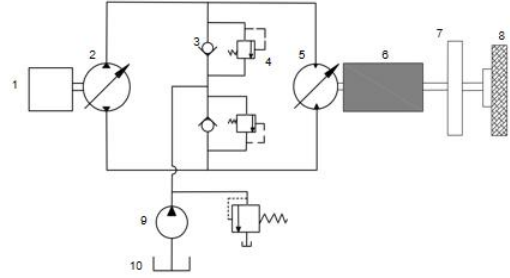

1.Diesel engine 2. Bi-way variable-displacement pump 3.Check valve 4.High pressure relief valve 5.Variable-displacement motor 6.Two-speed transmission 7.Axle 8.Driving wheel 9.Charge pump 10.Oil tank

Fig.1 HST system structure of forklift truck

## 1.1 Diesel Engine Model

This study aims to improve power matching in vehicle travel systems, emphasizing overall operational performance. A test-data modeling approach was used to construct the diesel engine model (the engine is Yu-Chai YCF36100-T480). Its universal characteristic curve is depicted in Fig.2

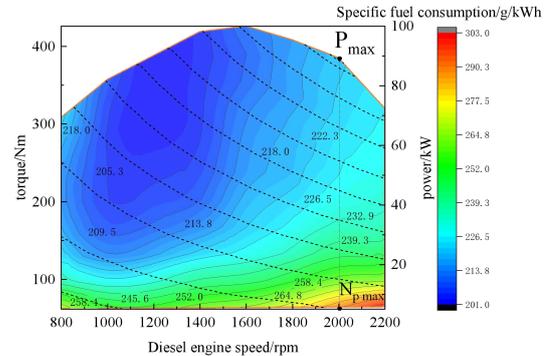

Fig.2 Diesel engine universal characteristic curve

To simulate the dynamic characteristics of the diesel engine, the output power is further expressed as a function of the power consumption of internal rotating components and auxiliary systems.

$$T_{ed} = T_e - J_e \cdot \dot{n}_e - T_a \quad (1)$$

where, $T_{ed}$ is the dynamic torque of the diesel engine $(N.m)$; $J_e$ is the inertia of the internal rotating parts of the diesel engine $(kg \cdot m^2)$; $T_a$ is load torque of the diesel engine auxiliary parts $(N.m)$.

## 1.2 Model of pump and motor

In closed-loop HST systems, the pump-motor system theoretically ideally satisfies flow conservation. However, practical operation requires comprehensive consideration of internal leakage, external leakage, and fluid compressibility. The leakage effects between the pump and motor can be accounted for within the volumetric efficiency parameters, while fluid compressibility impacts are neglected in this analysis. Thus, the governing equation is derived as follows:

$$Q_{p \cdot out} = Q_{m \cdot in} + C_M \cdot p_H \qquad (2)$$

Where, $Q_{p \cdot out}$ is output flow of the variable displacement pump (L/min), $Q_{m \cdot in}$ is input flow of the variable displacement motor (L/min), $C_M \cdot p_H$ is the relief valve overflow flow rate (L/min). Due to the dual-variable system in this study, the pump displacement and rotational speed can be dynamically adjusted in real-time according to the motor flow demand, significantly reducing or even eliminating overflow losses. Consequently, overflow flow effects will not be discussed in subsequent power matching analyses.

In the HST system, the diesel engine is rigidly connected directly to the variable-displacement pump. Therefore, the following equation is derived:

$$n_e = n_p \qquad (3)$$
$$T_{p \cdot in} = T_{ed} \qquad (4)$$

where, $n_p$ is the pump speed (r/min) $T_{p.in}$ is the input torque of pump(N.m) In HST system, output flow of variable-displacement pump is:

$$Q_{p \cdot out} = \frac{V_p \cdot n_p \cdot \eta_{pv}}{1000} \qquad (5)$$

Where,
$V_p$ is the actual displacement of the pump (mL/r), $\eta_{pv}$ is volumetric efficiency of pump.
The input torque of pump is expressed as $T_{p.in}$

$$T_{p \cdot in} = \frac{V_p \cdot \Delta p}{2\pi \eta_{pm}} \qquad (6)$$

where, $\Delta p$ is the pump inlet-outlet pressure differential between ($Mpa$); $\eta_{pm}$ is the pump mechanical efficiency. Input flow of motor is:

$$Q_{m \cdot in} = \frac{V_m \cdot n_m}{1000 \eta_{mv}} \qquad (7)$$

where, $V_m$ is the actual displacement of the motor (mL/r), ; $n_m$ is the motor speed(r/min); $\eta_{mv}$ is the volumetric efficiency of motor.
Output torque of motor is:

$$T_{m \cdot out} = \frac{V_m \cdot \Delta p \cdot \eta_{mm}}{2\pi} \qquad (8)$$

Where, $T_{m \cdot out}$ is output torque of motor ($N.m$); $\eta_{mm}$ is motor mechanical efficiency.

Pump and motor efficiency is affected by system pressure, rotational speed, displacement, and other factors. In this study benchtop testing was conducted on hydraulic pump and motor chosen for vehicle applications. Parameters such as flow rate, pressure, efficiency, and leakage were recorded and analyzed across various operating conditions. As demonstrated in Fig.3, the efficiency characteristics of both pump and motor exhibit distinct trends under different working states.

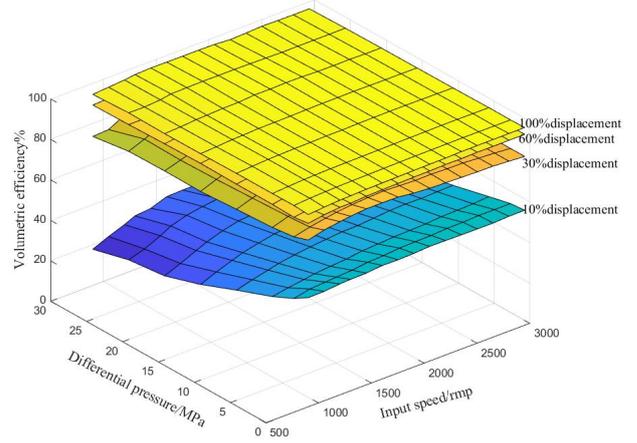

（a）Volumetric Efficiency of Pump

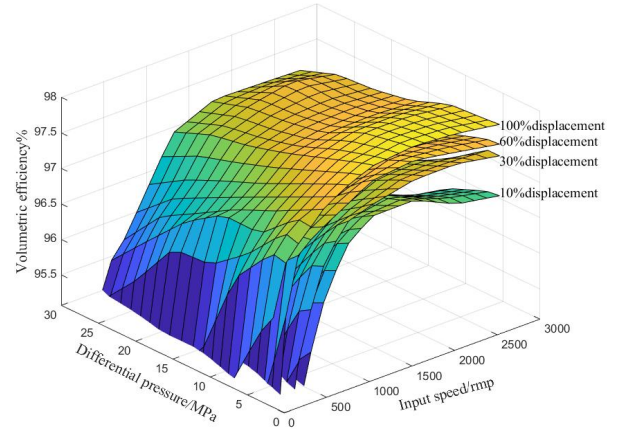

（b）Volumetric Efficiency of Motor

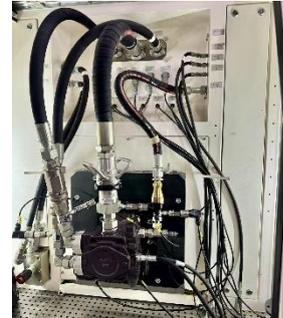

（c）Pump and Motor Test Bench
Fig.3 Efficiency model of pump and motor

## 1.3 Vehicle Model

The dynamics model of the vehicle is presented, concentrating exclusively on longitudinal performance

characteristics. The equilibrium equations, derived from the motion dynamics during driving, are as follows:

$$\Sigma F = mgf\cos\alpha + mg\sin\alpha + \delta m\frac{dv}{dt} + \frac{C_D A v^2}{21.15} \quad (9)$$

Where, $m$ is vehicle mass (kg), $g$ is gravitational acceleration(m/s²), $f$ is rolling resistance coefficient, α is slope angle (°), δ is rotation inertia conversion factor, $v$ is vehicle speed (km/h), $C_D$ is the aerodynamic drag coefficient, A is frontal area($m^2$)。

## 2 Power Matching Strategy Analysis
### 2.1 Control Target

The control strategy design prioritizes four key requirements: ①adaptability to variable vehicle loads, ②optimal diesel engine load point selection, ③driver demand satisfaction, ④pipeline flow and pressure loss minimization to enhance system efficiency. Based on these requirements, a control strategy for the HST system was developed through systematic investigation of the forklift truck's load condition characteristics.

### 2.2 Analysis of Vehicle Acceleration Process

In this study, the diesel engine controller operates in a pedal-to-speed governing mode, where the driver's pedal input directly correlates with the target engine speed. For instance, a 100% pedal opening corresponds to the target engine speed at $n_{p.max}$ in Fig.4. The diesel controller dynamically adjusts fuel delivery quantity based on vehicle load conditions to maintain the target engine speed. At the target speed, the engine's load ratio varies in real-time with load fluctuations.

Utilizing the engine's speed-torque curve external characteristic, the maximum allowable pump displacement $V_{p1}$ at a target rotational speed can be calculated. This is achieved by integrating Equations (2)(4)(6) and (8)

$$V_{p1} = \frac{2\pi T_e \eta_{pm}}{\Delta p} \quad (10)$$

Given a hydraulic pump displacement $V_{p1}$, pump speed $n_e$, and rated pressure differential ($\Delta p$), the maximum power that pump can output is calculated as:

$$P_{p1\cdot max} = \frac{\Delta p \cdot n_e \cdot V_{p1} \cdot \eta_{pv}}{1000} \quad (11)$$

With this output power value, motor can maintain its max displacement within speed range 0-$n_{m1}$, where $n_{m1}$ is:

$$n_{m1} = \frac{V_{p1} \cdot n_e \cdot \eta_{pv} \cdot \eta_{mv}}{V_{m\cdot max}} \quad (12)$$

where, $V_{m.max}$ is the max displacement of motor($mL/r$).

When the motor speed is below $n_{m1}$, motor operates within the constant torque output range, the motor can maintain its max displacement $V_{m.max}$ to achieve optimal acceleration. During this phase, pump displacement dynamically adjusts to match the motor's flow rate demand, following the relationship:

$$V_p = \frac{V_{m\cdot max} \cdot n_m}{n_{e1} \cdot \eta_{pv} \cdot \eta_{mv}} \quad (13)$$

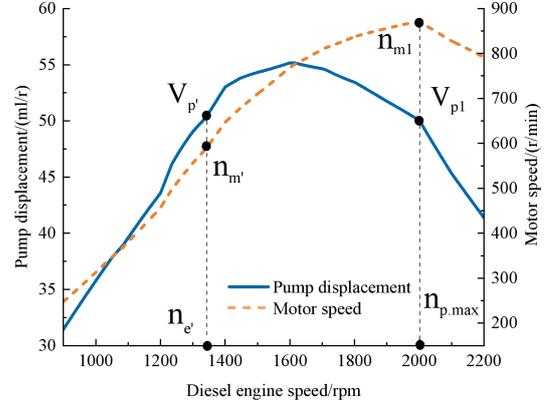

Fig.4 The relationship between diesel engine speed, pump displacement and motor speed

As motor speed reaches $n_{m1}$ it enter's the constant power region,Concurrently the pump displacement attains its max permissible displacement $V_{p1}$. at current engine speed $n_e$, while the engine achieves its full - load power at the current engine speed $n_e$. As vehicle speed increases further, the motor displacement adjusts along the constant power curve, during this procedure the motor displacement is:

$$V_m = \frac{V_{p1} \cdot n_{e1} \cdot \eta_{pv} \cdot \eta_{mv}}{n_m} \quad (14)$$

When the motor displacement reach its minimum value $V_{m.min}$, the motor speed is:

$$n_{m2} = \frac{V_{p1} \cdot n_{e1} \cdot \eta_{pv} \cdot \eta_{mv}}{V_{m\cdot min}} \quad (15)$$

When the moter speed reaches $n_{m2}$ if the driving force remains greater than the resistance, the vehicle speed can continue toncrease. Concurrently,the motor flow demand rises with increasing motor speed,leading to a"eduction in pump outlet pressure, Under this condition, the pump transitions into reverse drag operation mode, where its displacement is no longer constrained by the engine's externalcharacteristic power limits,In this operational regime, pump displacement adjustent prioritizesengine speed regulation, lf pump displacement remains at $V_{p1}$ the pump speed will graduallyincrease with vehicle speed until reaching the maximum allowable speed when the motor speed hits $n_{m3}$ To prevent overspeedinduced damage to components, the pump displacement should be incrementally adjusted to"educe speed, The pump displacement under such conditions is determined according to itsmaximum allowable speed $n_{e.max}$:

$$V_p = \frac{V_{m\cdot min} \cdot n_m}{n_{e\cdot max} \cdot \eta_{pv} \cdot \eta_{mv}} \quad (16)$$

Theoretical curves illustrating the relationship between pump/motor displacement and motor speed, derived from Equations (10)-(16), are shown in Fig.5.

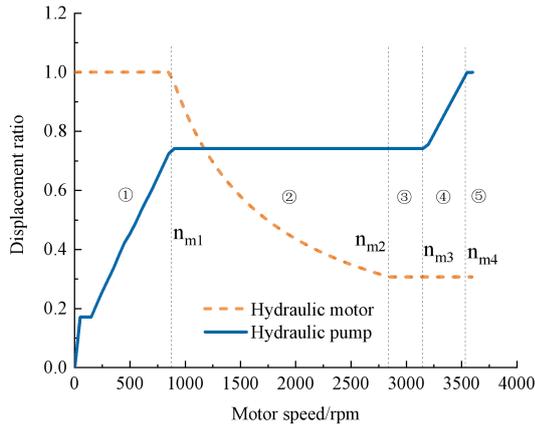

Fig.5 The displacement of pump and motor varies with the speed of motor

As can be seen in Fig.5, displacement control of the HST system can be divided into five distinct stages according to motor rotational speed:

Stage 1 ($n_m < n_{m1}$), the motor operates at its maximum displacement, while the pump displacement increases proportionally with the rising motor speed.

Stage 2 ($n_{m1} \leq n_m < n_{m2}$), the motor displacement decreases along the constant power cuve, whereas the pump displacement remains constant at $V_{p1}$.

Stage 3 ($n_{m2} \leq n_m < n_{m3}$), both the motor and pump displacements are maintained constant. The pump enters reverse drag operation mode, where pump displacement is no longer limited by diesel engine power limits.

Stage 4 ($n_{m3} \leq n_m < n_{max}$), the motor displacement is minimized, while the pump displacement reaches its allowable maximum limit. To prevent potential overspeed-induced pump damage, the pump displacement is actively adjusted to ensure that the speed does not exceed the permissible threshold.

Stage 5, the pump displacement may reach its maximum value while the motor displacement retains its min value. Both the pump and motor displacements keep constant during this stage.

The above analysis provides a detailed description of the vehicle acceleration process under a fixed pedal opening while system pressure difference maintaining at the rated value $\Delta p$, with the control objective set to maximize dynamic performance while maintaining the system pressure difference at the rated value. The acceleration process for different pedal openings can be analyzed in a similar method.

## 2.3 Optimal Power Matching Scheme

The diesel engine is controlled in a pedal-to-speed mode, where the engine load rate adjusts according to the load. To meet the demand for maximum acceleration capability, the diesel engine should operate at load rate 100%. If the HST system maintains the rated pressure difference $\Delta p$, the pump displacement should be adjusted according to the process described in Fig.5 at the target speed. Furthermore, the curve in Fig.5 can be converted into a displacement ratio and extended along the pedal opening dimension to form the target displacement ratio surface, as shown in Fig.6.

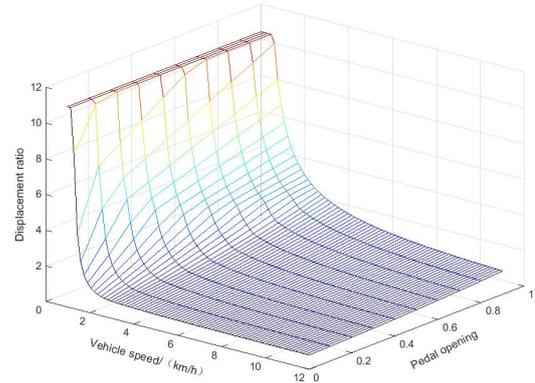

Fig.6 Target displacement ratio diagram (target displacement ratio = motor displacement/pump displacement)

## 3 Control strategy for HST system

### 3.1 Steady-State Control Strategy for HST System

The HST system utilizes a Hengli HP4VG-75-EP3 variable displacement pump and a Hengli M60V-115 variable displacement motor. Both the pump and motor employ electro-proportional displacement control. The calibration curves of the pump and motor displacement as a function of control current are illustrated in Fig.7. Based on the steady-state calibration characteristics, it can be observed that there is hysteresis both in the pump and motor current-displacement curve (Due to the fully symmetric structure of the pump's displacement control mechanism in forward and reverse directions, only the first-quadrant hysteresis is displayed in the figure). The presence of this hysteresis effect requires dynamic compensation of the target control current in the control strategy.

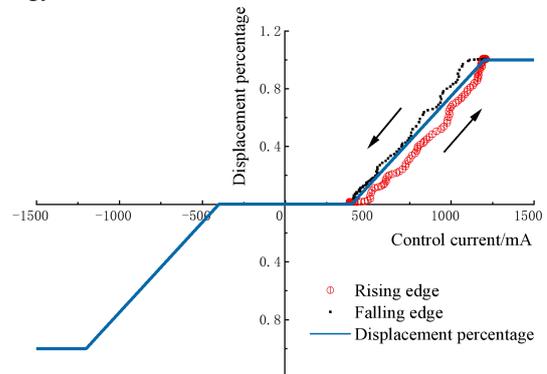

(a) Pump Current-Displacement Curve

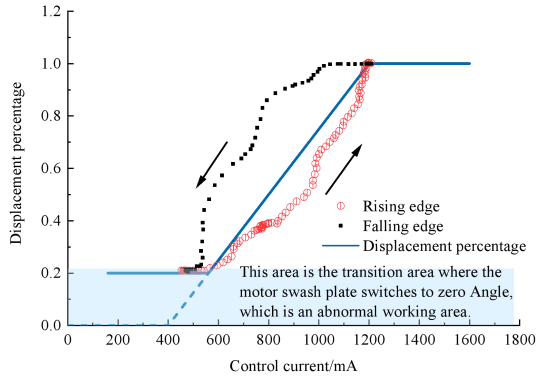

(b) Motor Current-Displacement Curve

Fig.7 The relationship between pump and motor displacement and control current and hysteresis effect

The control strategy determines the control electric current for the pump and motor separately based on the HST displacement ratio. The control block diagram is showed in Fig.8.

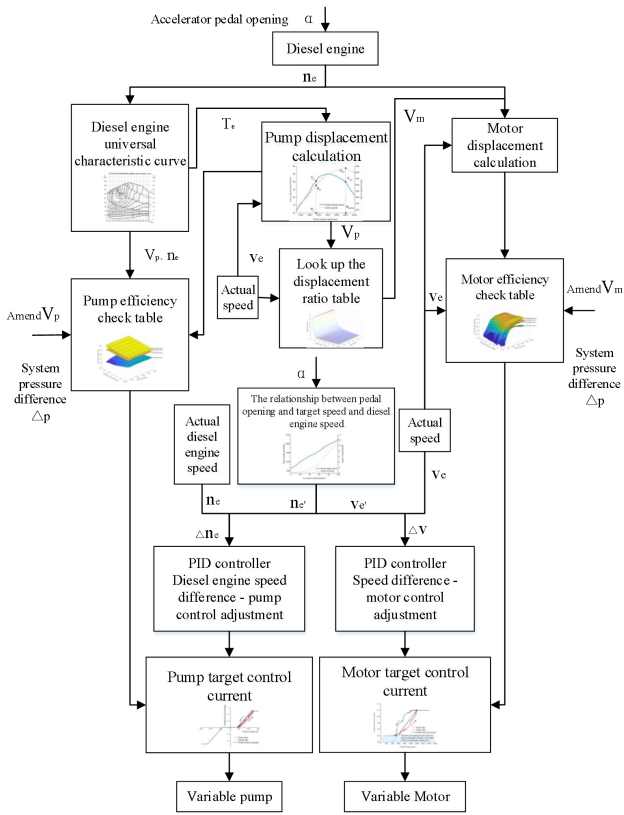

Fig.8 Hydraulic pump, hydraulic motor control strategy block diagram

The accelerator pedal opening corresponds to the target engine speed. By integrating the target speed, the engine's universal characteristic curve, the motor speed, and the pressure difference $\Delta p$, the corresponding motor displacement can be determined. Based on the electro-proportional calibration current-displacement curves of the pump and motor, the target control currents for both components can be obtained. The control currents are then adjusted using a PID method based on feedback from the vehicle's response state.

## 3.2 Dynamic compensation of Control Strategy

The steady-state control strategy, as mentioned above, focuses on power and load matching. Nevertheless, dynamic driving performance is equally crucial. To mitigate the risk of unintended drastic alterations in the vehicle's operational state due to significant variations in the target control current, while simultaneously fulfilling the rapid response demands during braking and emergency acceleration events, a dynamic compensation mechanism has been integrated. This enhancement is superimposed upon the foundational steady-state control objective, ensuring robust performance under both steady and dynamic driving conditions.

3.2.1 Diesel engine overload dynamic compensation

Predict overload risk of the engine and develop a dynamic compensation strategy for pump displacement control to prevent engine stalling, as shown in Figure 9.

3.2.1 Dynamic Compensation for Diesel Engine Overload

To anticipate potential diesel engine overload and prevent engine stalling, a dynamic compensation strategy for pump displacement has been developed, as illustrated in Fig.9.

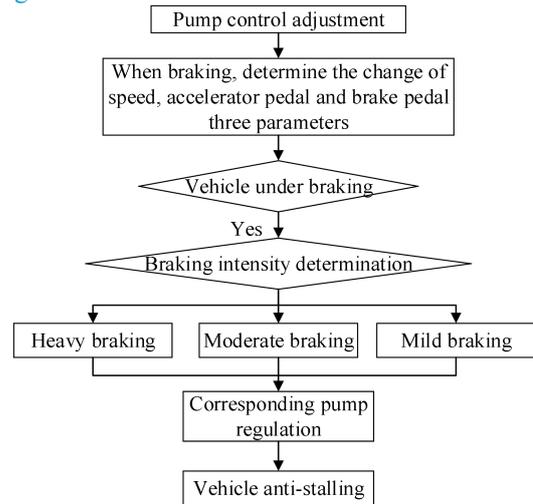

Fig.9 Pump displacement dynamic compensation strategy block diagram

3.2.2 Emergency braking dynamic compensation

The purpose of this dynamic compensation strategy is to adjust the motor displacement during emergency braking to cope with load shock and rapidly changes in vehicle speed. The strategy is shown in Fig.10.

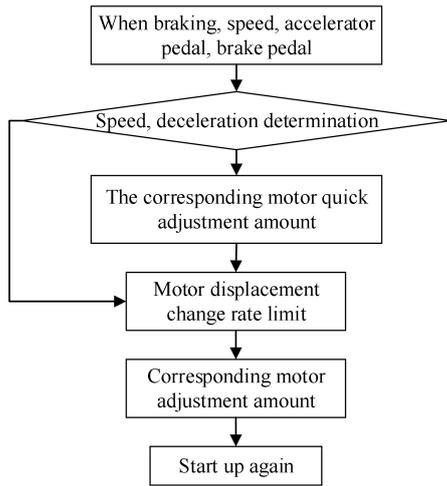

Fig.10 Motor displacement dynamic compensation strategy block diagram

### 3.2.3 Driver's Expectation Emergency Judgment and Compensation

The target displacement ratio in Fig.6 aims to achieve maximum acceleration objectives. However, for practical applications, it is essential to assess the operational scenario and interpret driver intention by taking into account both the pedal position and its change rate. Consequently, adaptive strategies are formulated accordingly, as depicted in Fig.11.

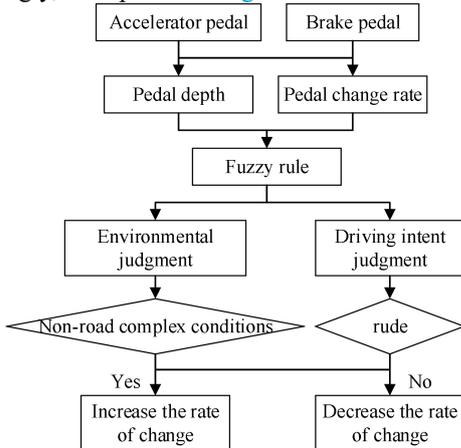

Fig.11 The driver expects to adjust the block diagram

## 4 Test Verification

The control strategy for the HST system proposed in the preceding sections was validated through on-road vehicle tests. The test setup is described as follows.

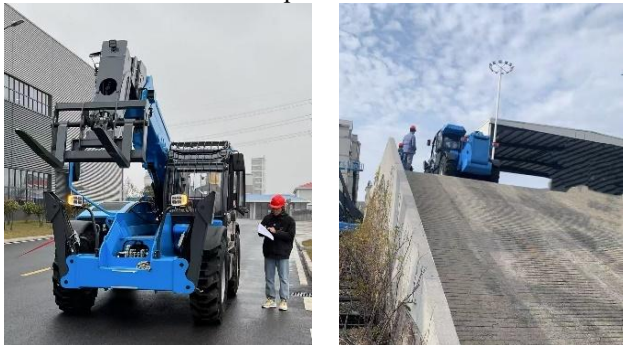

Fig.12 Test vehicle

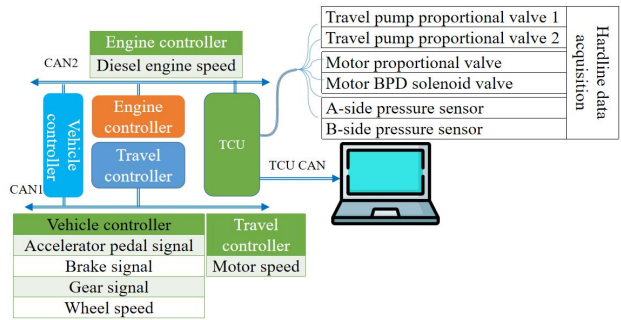

Fig.13 Signal acquisition system in vehicle test

The test vehicle shown in Fig.12 was used to conduct test on the HST system. Control parameters and vehicle response data were collected, with signals acquired via both CAN bus and hardwired connections, as depicted in Fig.13.

### 4.1 **Uphill operating condition**

The vehicle was positioned stationary at the origin line on a flat asphalt pavement, facing the 40% gradient slope. The pedal opening was actuated to its predefined position and maintained until the rear wheels achieved slope top.

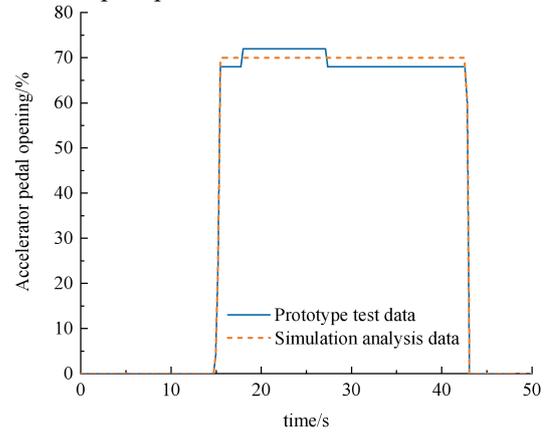

（a） Pedal opening

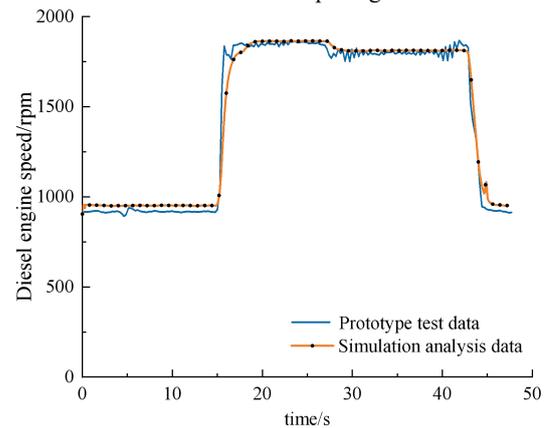

（b） Engine speed

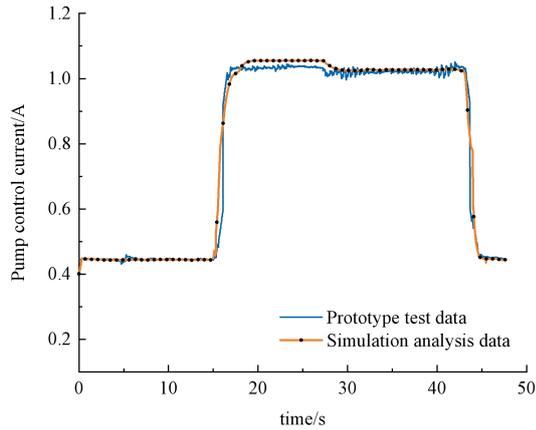

(c) Current of pump

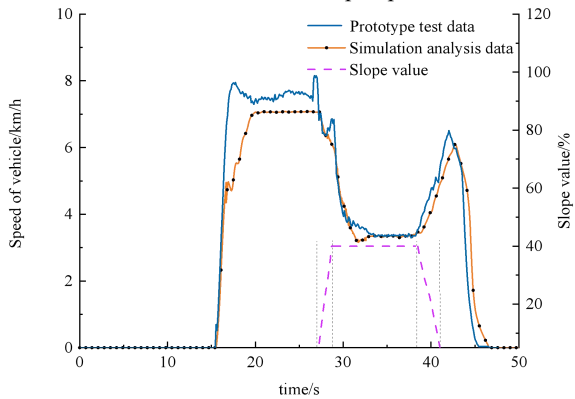

(d) Vehicle speed

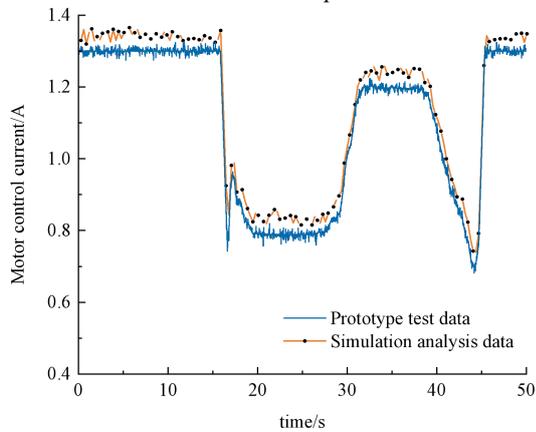

(e) Current of motor

Fig.14 Comparison of vehicle uphill condition test

In Fig.14, the vehicle state changes as follows:

① Under engine idle condition, the pump displacement remains at a lower level while the motor displacement is maintained at a higher level (ready to provide high torque for vehicle starting).

②At 15th second, the accelerator pedal opening increases to 70%, and the engine target speed correspondingly rises to the value associated with a 70% pedal opening. As the vehicle accelerates, the motor displacement decreases rapidly while the pump displacement increases accordingly. This coordinated adjustment achieves flow matching and load adaptation until the vehicle speed stabilizes.

③Starting from 27th second, the front wheels enter a 40% gradient slope, as shown in Fig.14 (d), with the rear wheels entering the slope at 29th second. The increased slope resistance causes the vehicle speed to decrease, prompting an adaptive adjustment of the motor displacement. Eventually, the vehicle maintains a stable speed while climbing the 40% gradient slope.

④ At 38th second, the front wheels reach the top of the slope, and the reduced resistance due to the change in slope causes the vehicle speed to increase slightly, accompanied by a reduction in motor displacement. At 41th second, the rear wheels reach the slope top, and the vehicle transitions to zero-gradient pavement driving conditions.

The dynamic evolution of control variables during the uphill traversal demonstrates the operational mechanism of the proposed control strategy in Section 2. This test validation confirms that the developed control strategy satisfies the performance requirements for vehicle uphill operation, achieving gradeability of 40% at 3 km/h with 70% pedal opening；

### 4.2 Hill-start test

The vehicle was initially positioned statically on a 40% gradient slope with the brake pedal depressed. Upon release of the brake pedal, the acceleration pedal was promptly pressed to a 70% opening and maintained at this position until the vehicle reached the slope top.

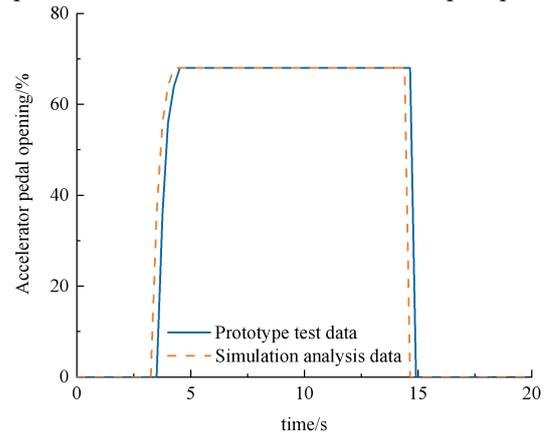

(a) Pedal opening

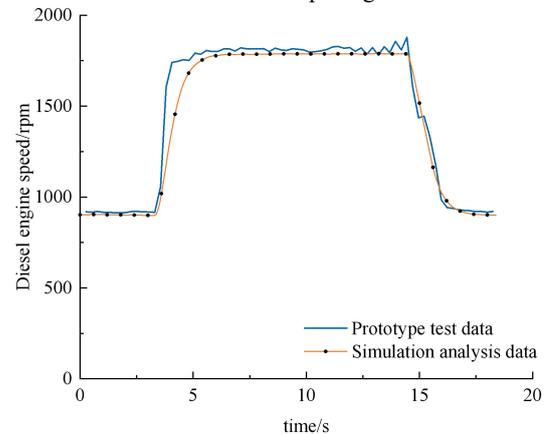

(b) Speed of engine

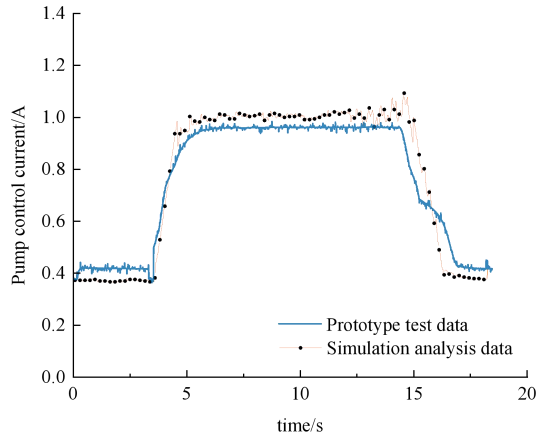

（c）Current of pump

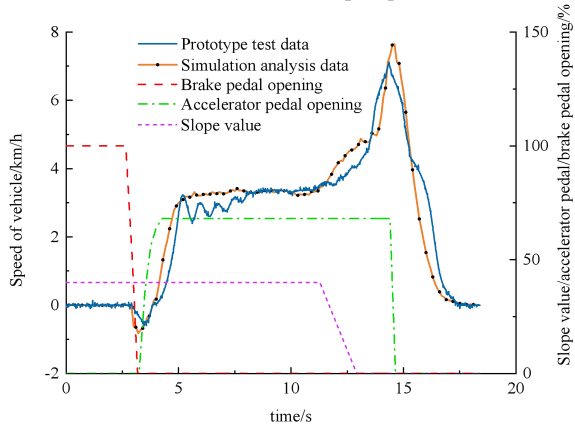

（d）Vehicle speed

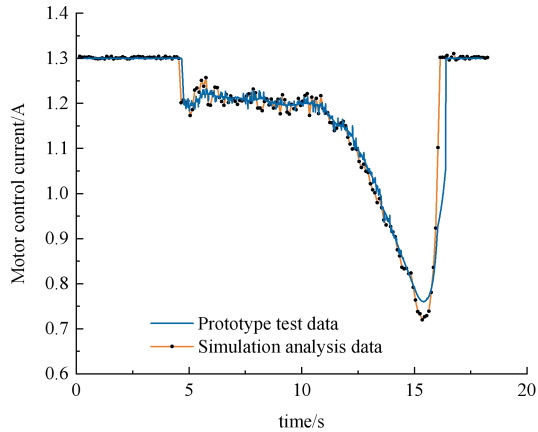

（e）Current of motor

Fig.15 Comparison of vehicle ramp start test

As illustrated in Fig.15, the vehicle's dynamic response and control parameters evolution demonstrate the efficacy of the proposed control strategy.

① The brake pedal was depressed and the engine operated in idle condition. The pump control current remains at a lower displacement, while the motor displacement is maintained at a higher level (ready to provide high torque for vehicle starting).

② At 3rd second, the brake pedal was released, causing a rollback of the vehicle. The accelerator pedal was then pressed to 70%. Due to the large slope resistance, the motor remained at its maximum displacement until the vehicle reached the top of the slope.

③ At 11th second, the front wheels reached the top of the slope, and at 13th second, the rear wheels reached the slope top, where the slope resistance decreased. The vehicle speed increased rapidly, and the motor displacement was quickly reduced. By 15th second, the accelerator pedal opening decreased to 0%.

The dynamic evolution of control variables during the hill-start traversal demonstrates the operational mechanism of the proposed control strategy in Section 2. This test validation confirms that the developed control strategy satisfies the performance requirements for vehicle hill-start operation, achieving hill-start gradeability of 40% with 70% pedal opening.

### 4.3 Brake condition

Under braking conditions, additional dynamic compensations strategies were incorporated based on the control strategies outlined in Fig.5 and Fig.6.

The vehicle was positioned stationary at the origin line on a flat asphalt pavement and accelerated to its maximum speed with a 100% accelerator pedal opening. After maintaining this speed for a certain duration, the brake pedal was pressed down until the vehicle came to standstill. The data recording began once the vehicle reached a stable speed during acceleration. Two braking scenarios were tested: light braking (brake pedal reaching 40% depth within 1 second) and heavy braking (brake pedal reaching 100% depth within 0.1 seconds).

The braking process proceeded as follows:

① At 6th second, the accelerator pedal was released, causing the pump's target speed to rapidly decrease. Correspondingly, the target pump displacement also declined, resulting in the pump's control current swiftly reducing as the pedal opening decreased.

② As the brake pedal was pressed, the pump entered reverse drag operation mode. Under heavy braking conditions, the pump's control current was sharply reduced, with an overshoot adjustment employed to accelerate the reduction of pump displacement. This is illustrated in Fig.16(c), reflecting the dynamic compensation described in Section 3.2 of the paper.

③ Under light braking conditions, the control currents for both the pump and motor adjusted gradually. Toward the end of braking, the motor control current exhibited a step increase to its maximum value, preparing for next vehicle restart.

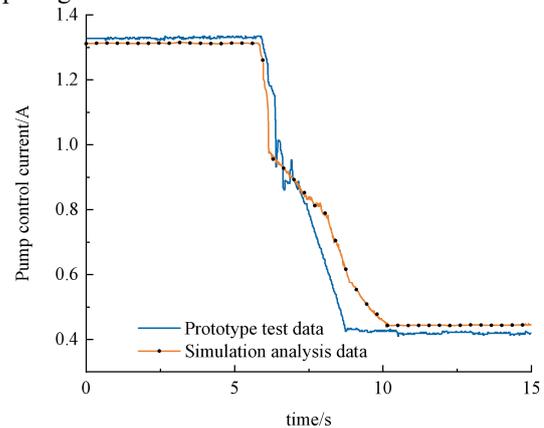

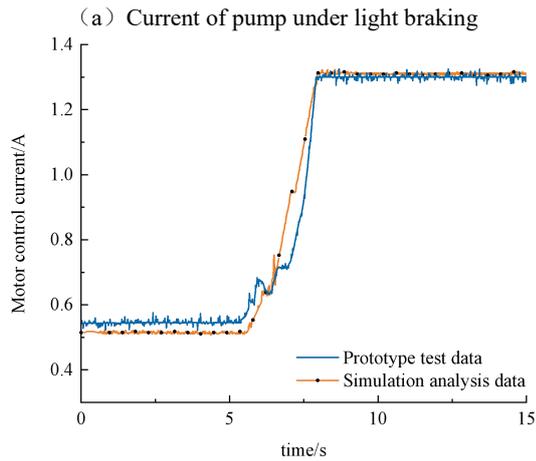

（a）Current of pump under light braking

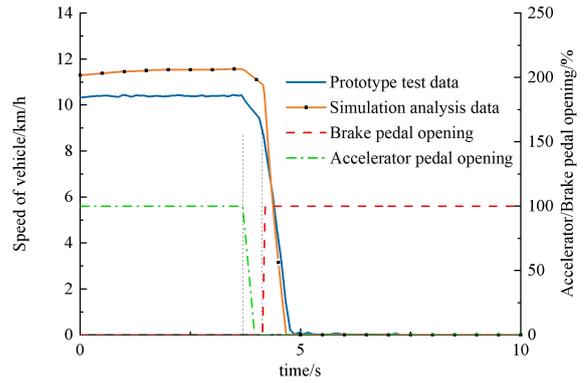

（f）Vehicle speed and pedal opening under heavy braking
Fig.16 Vehicle braking condition test

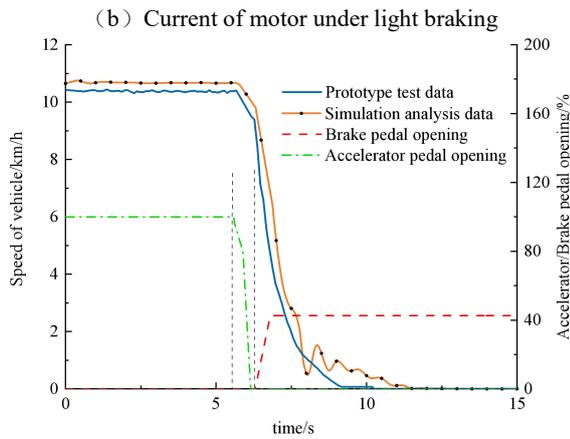

（b）Current of motor under light braking

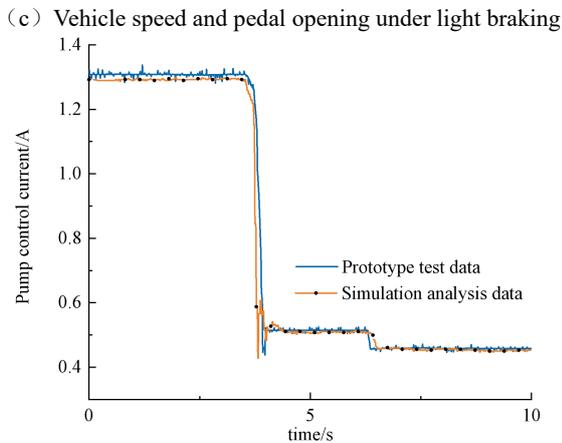

（c）Vehicle speed and pedal opening under light braking

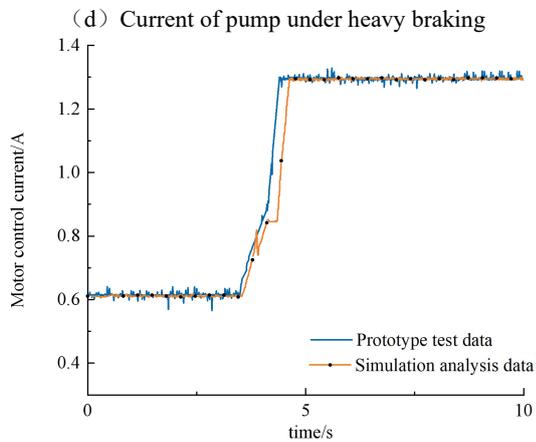

（d）Current of pump under heavy braking

（e）Current of motor under heavy braking

## 5 Analysis and Conclusion

This paper developed a dynamic model of a dual-variable HST system and its associated vehicle, analyzed the vehicle's acceleration process. With the goal of maximizing acceleration, an innovative pump-motor displacement ratio matching strategy was proposed. Based on this foundation, an integrated control framework is constructed, incorporating engine overload emergency protection mechanism, a driver intention dynamic compensation algorithm, and a braking condition control compensation strategy. This framework effectively balances power matching strategy with rapidity and stability of dynamic response.

Through comparative validation using virtual simulation and real-vehicle testing, the control strategy proposed in this study for the forklift traveling system demonstrates excellent real-time response performance. The strategy dynamically adjusts control parameters based on changes in vehicle state during uphill driving, zero-gradient pavement cruising, and braking conditions, fully meeting the performance requirements of the forklift traveling system.

# References

[1] A controlled diesel drive line with hydrostatic transmission: Part 2 - dynamic properties at periodic loading. International Journal of Vehicle Design [Internet]. [cited 2025 Apr 11]; Available from: https://www.indersciencenonline.com/doi/10.1504/IJVD.2005.007287

[2] Mapping the Efficiency for a Hydrostatic Transmission | J. Dyn. Sys., Meas., Control. | ASME Digital Collection [Internet]. [cited 2025 Apr 11]. Available from: https://vpn.jlu.edu.cn/https/4469646964613131323744696461ab6bf16d07c1ac1ed85aeb6fd86999f7881bd03bb0ec37d7efede4225842/dynamicsystems/article/138/3/031004/371218/Mapping-the-Efficiency-for-a-Hydrostatic

[3] Research on the New Hydrostatic Transmission System of Wheel Loaders Based on Fuzzy Sliding Mode Control [Internet]. [cited 2025 Apr 10]. Available from: https://www.mdpi.com/1996-1073/17/3/565

[4] Kumar N, Kumar R, Sarkar BK, Maity S. Condition monitoring of hydraulic transmission system with variable displacement axial piston pump and fixed displacement motor. Materials Today: Proceedings. 2021;46:9758–65.

[5] Chen H, Wang M, Ni X, Cai W, Zhong C, Ye H, et al. Design of Hydrostatic Power Shift Compound Drive System for Cotton Picker Experiment. Agriculture. 2023 Aug 10;13(8):1591.

[6] Decentralized multivariable modeling and control of wind turbine with hydrostatic drive-train-All Databases [Internet]. [cited 2025 Apr 11]. Available from: https://vpn.jlu.edu.cn/https/4469646964613131323744696461bd7dfe6705dba81ec955e469996699f3a2c9e8c2ae15ee71b2/wos/alldb/full-record/PQDT:68788961

[7] Optimal Control and Architecture Design Optimization for Hydraulic Drive Train of a Compact Track Loader-All Databases [Internet]. [cited 2025 Apr 11]. Available from: https://vpn.jlu.edu.cn/https/4469646964613131323744696461bd7dfe6705dba81ec955e469996699f3a2c9e8c2ae15ee71b2/wos/alldb/full-record/WOS:000702263300100

[8] 孙恬恬, 毛恩荣, 傅梁起, 宋正河, 李臻, 李平. 基于变量马达控制的喷雾机驱动防滑系统设计与试验. 农业机械学报. 2024;55(5):158–66.

[9] Soma A, Bruzzese F, Mocera F, Viglietti E. Hybridization Factor and Performance of Hybrid Electric Telehandler Vehicle. IEEE Trans on Ind Applicat. 2016 Nov;52(6):5130–8.

[10] Puras B, Raush G, Freire J, Filippini G, Roquet P, Tirado M, et al. Development of a Virtual Telehandler Model Using a Bond Graph. Machines. 2024 Dec 4;12(12):878.

[11] 赵春江, 魏传省, 付卫强, 尚业华, 张光强, 丛岳. 静液压传动拖拉机定速巡航控制系统设计与试验. 农业机械学报. 2021;52(4):359–65.

[12] Cui HX, Feng K, Li HL, Han JH. Response Characteristics Analysis and Optimization Design of Load Sensing Variable Pump. Mathematical Problems in Engineering. 2016 Jan;2016(1):6379121.

[13] Wan L rong, Lu Y jie, Zeng Q liang, Gao K dong, Jiang S bo. The Research on Comprehensive Performance Evaluation of Axial Piston Pump Based on AHP. Mathematical Problems in Engineering. 2018 Jan;2018(1):9469064.

[14] Parlapanis C, Müller D, Frontull M, Sawodny O. Modeling of the Work Functionality of a Hydraulically Actuated Telescopic Handler. IFAC-PapersOnLine. 2022;55(20):253–8.

[15] Caffaro F, Cremasco MM, Preti C, Cavallo E. Ergonomic analysis of the effects of a telehandler's active suspended cab on whole body vibrat

ion level and operator comfort. International Journal of Industrial Ergonomics. 2016 May;53:19–26.

[16] Campana D. Analysis of algorithms for autonomous driving of a telescopic handler [Internet] [laurea]. Politecnico di Torino; 2023 [cited 2025 Apr 10]. Available from: https://webthesis.biblio.polito.it/26642/

[17] 孙景彬, 楚国评, 潘冠廷, 孟宏, 刘志杰, 杨福增. 遥控全向调平山地履带拖拉机设计与性能试验. 农业机械学报. 2021;52(5):358–69.

[18] 李胜, 宋大凤, 曾小华, 贺辉, 聂利卫, 王继新. 重型卡车轮毂马达液压驱动系统建模与仿真. 农业机械学报. 2012;43(4):10–4.

[19] 孔祥东, 宋豫, 艾超. 变转速输入定量泵-恒转速输出变量马达系统恒转速控制方法研究. 机械工程学报. 2016;52(8):179–90.

[20] Kumar N, Dasgupta K, Ghoshal S. Dynamic analysis of a closed-circuit hydrostatic summation drive using bent axis motors. Proceedings of the Institution of Mechanical Engineers, Part I: Journal of Systems and Control Engineering. 2015 Sept;229(8):761–77.

[21] 汪浒江, 王涛, 林宇, 杜甫, 冯文韬. 浮动式配流盘柱塞泵马达系统动力学建模及特性研究. 汽车工程. 2024;46(11):2122–32.

[22] 程旭, 彭增雄, 荆崇波. 拖拉机液压机械无级传动系统液压泵马达全工况效率模型研究. 北京理工大学学报. 2025;45(2):154–64.

[23] 朱从民. 静液压传动车辆的复合控制. 农业机械学报. 2005;(4):26–9.

[24] 朱晨辉, 王淼森, 陈博, 张红梅, 朱骐, 何勋, et al. 丘陵区履带式烟叶采收机液压行驶系统的创新设计与试验. 河南农业大学学报. 2022;56(4):532–42.

[25] 朱镇, 蔡英凤, 陈龙, 夏长高, 田翔, 施德华. 多功能动力传动装置设计方案研究. 汽车工程. 2020;42(10):1378-1385+1403.